\documentclass{article}
\usepackage{spconf,amsmath,graphicx,hyperref}

\usepackage{multirow}
\usepackage{array}
\usepackage{xspace}
\usepackage{booktabs}
\usepackage{siunitx}
\usepackage{algorithm}
\usepackage{algpseudocode}
\usepackage{balance}
\usepackage{amssymb}   
\usepackage{xcolor}
\usepackage{enumitem} 
\definecolor{prompt}{HTML}{6B7280}     
\definecolor{speech}{HTML}{2563EB}     
\definecolor{noiseA}{HTML}{0D9488}     
\definecolor{noiseT}{HTML}{DB2777}     
\definecolor{textout}{HTML}{EA580C}    
\definecolor{softp}{HTML}{8B5CF6}      

\title{Text-only adaptation in LLM-based ASR through text denoising}
\name{
 \begin{tabular}{c}
 Andr\'es Carofilis$^{1,\dagger}$, Sergio Burdisso$^{1,\clubsuit}$, Esa\'u Villatoro-Tello$^{1,\ddagger}$,  Shashi Kumar$^{1,2,\star}$, Kadri Hacioğlu$^{3,\star}$ \\
 Srikanth  Madikeri$^{4,\star}$, Pradeep Rangappa$^{1,\star}$, Manjunath K E$^{3,\star}$, Petr Motlicek$^{1,5,\star}$, \\
 Shankar Venkatesan$^{3,\star}$, Andreas Stolcke$^{3,\star}$\\
 \vspace{-2em}
 \end{tabular}
}
\address{$^{1}$ Idiap Research Institute \quad
$^{2}$ EPFL, Lausanne \quad
$^{3}$ Uniphore \quad
$^{4}$ Univ.~of Zurich \quad
$^{5}$ BUT, Brno \quad
}
\begin{document}
\ninept
\maketitle
%
\noindent\fcolorbox{red}{red!5}{%
  \parbox{\dimexpr\linewidth-2\fboxsep-2\fboxrule\relax}{%
    \footnotesize\color{red}%
    \textbf{NOTICE: this version has been superseded by a revised version.}
    In a subsequent internal reproduction of this work, errors were identified in the experimental pipeline and in the preparation of the evaluation data:
    the development and test partitions used here were incomplete and did not cover the full evaluation sets.
    The pipeline was reimplemented, the dataset partitions were corrected to include the complete development and test sets, and all experiments were rerun. As a result, the dataset statistics, results, and conclusions reported here have been revised.
    The revised study also releases the implementation and scripts required to reproduce them.
    The author list of the revised version reflects the contributions to that study.
    \textbf{Readers are encouraged to refer to the revised version, published at \mbox{INTERSPEECH~2026}:} \textcolor{blue}{\url{https://www.isca-archive.org/interspeech_2026/burdisso26_interspeech.html}}
  }%
}
\vspace{0.6em}
\begin{abstract} 

Adapting large language model (LLM)-based automatic speech recognition (ASR) systems to new domains using text-only data is a significant yet underexplored challenge. Standard fine-tuning of the LLM on the target domain text often disrupts the critical alignment between the speech and text modality learned by the projector, degrading performance. We introduce a novel text-only adaptation method that frames this process as a text denoising task. Our approach trains the LLM to recover clean transcripts from noisy inputs. This process effectively adapts the model to a target domain while preserving cross-modal alignment. Our solution is lightweight, requiring no architectural changes or additional parameters. Extensive evaluation on two datasets demonstrates up to 22.1\% relative improvement, outperforming recent state-of-the-art text-only adaptation methods.
\end{abstract}
\renewcommand*{\thefootnote}{\fnsymbol{footnote}}
\section{Introduction}\footnotetext{
\begin{tabular}{@{}lp{0.85\linewidth}}
$^\dagger$   & Performed experimentation and data preparation.\\
$^\clubsuit$  & Conceived the original idea and helped plan the experiments.\\
$^\ddagger$ & Involved in planning and supervised the work, writing the manuscript with support from $^\ddagger$ and $^\dagger$.\\
$^\star$ & Provided critical feedback and helped shape the research.
\end{tabular}
}
\renewcommand*{\thefootnote}{\arabic{footnote}}
\label{sec:intro}

Recently, there has been growing interest in integrating speech capabilities into large language models (LLMs) to enable seamless voice interaction and advance voice-driven applications, assistive technologies, and conversational AI more broadly ~\cite{goel2025audio,chu2024qwen2,ma2024embarrassingly,wang2023slm,huang2024multilingual, li2023prompting, ma2023can, yang2023generative, KHEDDAR2024102422, zhang2023speechgpt, das2024speechverse}. 


In this context, LLM-based ASR systems have emerged as a practical and computationally efficient alternative, focusing on transcription by leveraging a fixed, manually defined prompt during both training and inference~\cite{ma2024embarrassingly, wang2023slm,10445874, kumar2024performance, yang2025bridgingmodalitygapsoftly, DBLP:journals/corr/abs-2506-05671, sedlavcek2025approaching, yang24f_interspeech}. 
This fixed-prompt setup ensures consistency between the training objective and inference behavior, making it well-suited for applications where high-accuracy transcription is the primary goal. A key advantage of LLM-based ASR is the ease of combining strong pretrained speech encoders with powerful LLMs through a learnable projection layer, thereby leveraging advances from pretraining in both speech and text modalities. This modular design enables scalable, high-performance transcription without the need for costly instruction tuning. Intuitively, the projection layer learns to map speech representations into the text embedding space of the LLM (i.e., learns a speech-to-text alignment). Once projected, the resulting representation can be interpreted by the LLM as a noisy text, which the model reconstructs into a clean transcription through its inherent denoising capability. 

Despite these advantages, the training of LLM-based ASR typically relies on large amounts of paired audio–text data, which can limit scalability to new domains. In practice, such resources are often scarce or expensive to collect and transcribe.
Moreover, existing studies have indicated that performance may degrade when models are applied to domains that differ from the training data~\cite{kumar2024performance}, highlighting the importance of effective adaptation strategies.
Compared to collecting additional audio–text pairs, text-only adaptation offers a more practical alternative given the wide availability of text data.

Few studies have explored fine-tuning LLM-based ASR with unpaired text data while preserving cross-modal alignment between the speech projector and the LLM. Fang et al.~\cite{DBLP:journals/corr/abs-2506-05671} proposed using a monitoring metric to maintain alignment, but excessive text-only fine-tuning can still degrade recognition, and mitigation strategies only partially address this issue. Ma et al.~\cite{DBLP:conf/slt/MaLK24} introduced a two-step approach using trainable soft prompts as pseudo audio embeddings: first optimizing domain-specific soft prompts, then performing text adaptation with the soft prompts fixed. While effective, this method requires tuning additional hyperparameters, such as the number, initialization, and placement of soft tokens.

To address these challenges, we propose a novel text-only adaptation strategy that fine-tunes the LLM within an LLM-based ASR architecture by means of formulating the problem as a \textit{denoising task}. 
Our contributions are as follows: \textit{(i)} We reformulate text-only adaptation as a denoising task, training the LLM to reconstruct inputs that mimic the outputs of a speech projector in LLM-based ASR architectures. \textit{(ii)} We propose a lightweight training approach that simply consists of a multi-view noise-driven batching strategy, not requiring additional learnable parameters. \textit{(iii)} We present a thorough evaluation across two datasets, achieving up to 22.1\% relative improvement, surpassing the state-of-the-art.

\begin{figure}[t]
\centering
\includegraphics[width=\linewidth]{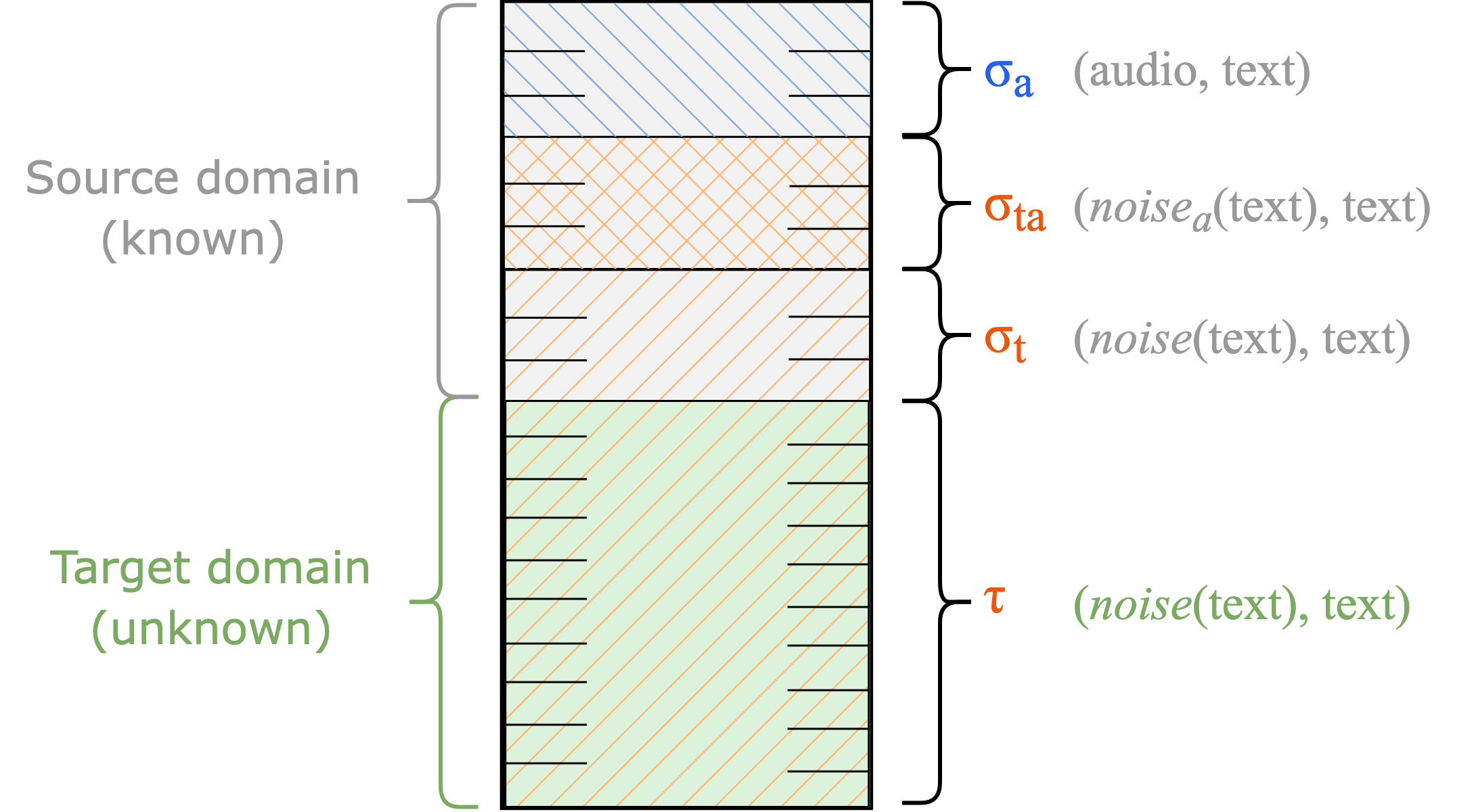}
\vspace{-0.4cm}
\caption{Batch composition used for fine-tuning the LLM during text-only adaptation to a target domain. Here, $\sigma$ and $\boldsymbol\tau$ denote the proportions of the batch that are drawn from the source domain ($\mathcal{D}{src}$) and the target domain ($\mathcal{D}{tgt}$), respectively.}
\label{figure:banking_defai_val_ppl}
\end{figure}


\section{Method}
\label{sec:method}

LLM-based ASR systems consist of three main components: (i)~a pretrained speech encoder  (e.g., Whisper~\cite{radford2023robust}, Hubert~\cite{hsu2021hubert}, or WavLM~\cite{chen2022wavlm}, etc.) that extracts frame-level acoustic representations, (ii)~a learnable projector (e.g., pooling-based~\cite{chu2024qwen2},CNN-based~\cite{das2024speechverse, Rajaa_SpeechLLM_Multi-Modal_LLM}, or linear-based~\cite{ma2024embarrassingly,10800077, kumar2024performance}) that maps these representations into the LLM input embedding space,\footnote{Also known as ``speech projector'' in the literature, with some referring to it as a ``speech adapter'' or ``connector''.} and (iii)~a pre-trained LLM (e.g., Llama~\cite{dubey2024llama}, Vicuna~\cite{vicuna2023} ) that acts as a decoder, and generates final transcripts. 

Prior work has shown that even a simple projector (two linear layers with a nonlinearity) combined with a frozen LLM is sufficient to obtain strong transcription performance~\cite{ma2024embarrassingly,kumar2024performance, yang2025bridgingmodalitygapsoftly}. This suggests that the projector learns to convert speech into a sequence of \emph{soft tokens} that approximate entries in the LLM vocabulary. For example, the projector may map the audio utterance ``yes that would be'' into embeddings resembling ``mmy Z \textbf{YesssS} S SGS \textbf{that Will} B \textbf{be} S S'', as reported in prior work~\cite{kumar2024performance, yang2025bridgingmodalitygapsoftly}. This illustrates how the projector outputs resemble a noisy transcript rather than raw speech features. As a result, the LLM is forced to interpret these inputs as a corrupted or noisy version of the transcript, which it then recovers.

We interpret this behavior as evidence that LLM-based ASR can be viewed as a \emph{denoising task}: the LLM learns to reconstruct the clean transcript from the noisy, text-like sequence produced by the projector. Building on this insight, we propose a novel approach to adapt LLM-based ASR models using only text data by explicitly teaching the LLM to denoise distorted transcripts from the target domain, even when no target-domain audio is available.

\subsection{Task Formulation}

Let $\mathcal{D}_{src} = \{(a_i, t_i)\}$ denote the source-domain dataset with paired audio $a_i$ and transcript $t_i$, and let $\mathcal{D}_{tgt} = \{t_j\}$ denote the target-domain dataset with transcripts only.\footnote{It is worth mentioning that in all our experiments, the text-only data consists of (ground truth) transcripts from conversational speech, rather than arbitrary text such as that found on web pages.} 
While standard LLM-based ASR training requires large numbers of $(a, t)$ pairs to achieve high recognition performance, we enable text-only adaptation by introducing a noise function $noise(\cdot)$. This function takes as input transcripts and generates perturbed variants that approximate the outputs of a trained projector in the LLM-based architecture.

Thus, given only text $t \in \mathcal{D}_{tgt}$, we replace the missing audio with $noise(t)$ and fine-tune the LLM to recover $t$. Adaptation is therefore reframed as training on $(noise(t), t)$ pairs, i.e., learning to solve a text denoising problem.

\subsection{Batch Construction for Text Denoising Adaptation}
\label{subsec:batch_construction}

Directly fine-tuning the LLM with $(noise(t), t)$ pairs from $\mathcal{D}_{tgt}$ leads to \emph{catastrophic forgetting}, i.e., the alignment between the speech encoder and the LLM degrades, and the projector’s mapping becomes ineffective, negatively impacting the model's performance. A similar observation has also been recently reported in ~\cite{DBLP:journals/corr/abs-2506-05671}. 
To mitigate this, we propose a more effective training batch composition for fine-tuning the LLM. Specifically, each batch is constructed as a mixture of examples from both source and target domains (see Figure~\ref{figure:banking_defai_val_ppl}). Let $\sigma_a$, $\sigma_{ta}$, $\sigma_t$, and $\tau_t$ denote the proportions (relative shares) of the following components in each batch:

\begin{itemize} [leftmargin=*, itemsep=0.2ex, topsep=0.4ex, parsep=0pt, partopsep=0pt]
    \item \textcolor{speech}{$\sigma_a$}: $(a, t)$ pairs from $\mathcal{D}_{src}$, i.e., paired audio and transcripts, to preserve the original speech–text alignment.
    \item \textcolor{textout}{$\sigma_{ta}$}: $(noise_a(t), t)$ pairs where $(a, t)\in\mathcal{D}_{src}$ and $noise_a(t)$ is obtained by projecting $a$ through the model’s projector and mapping the resulting embeddings to their nearest tokens in the LLM vocabulary. This approximates an optimal noise function, as it corresponds to projector-induced text noise.
    \item \textcolor{textout}{$\sigma_t$}: $(noise(t), t)$ pairs where $t$ is taken from $\mathcal{D}_{src}$ and $noise(t)$ is generated through random character substitutions and duplications. This serves as a naive approximation of $noise_a(t)$, obtainable without access to audio. By including both projector-induced and naive noise in the same batch for the source domain, the LLM is encouraged to learn to bridge the three views of $t$ (audio, $noise_a(t)$, and $noise(t)$).
    \item \textcolor{textout}{$\tau_t$}: $(noise(t), t)$ pairs from $\mathcal{D}_{tgt}$, analogous to the previous item but applied to the target domain, thereby driving adaptation by exposing the LLM to target-domain textual data.
\end{itemize}


The proportions are set so that $\sigma_a + \sigma_{ta} + \sigma_t + \tau = 1$. In practice, maintaining a small but nonzero $\sigma_a$ is critical to avoid forgetting the speech-to-text alignment between, while $\tau$ controls the strength of adaptation. Ideally, $\tau$ should be optimized on held-out validation data for each application setting, since it directly governs the balance between source retention and target specialization. In this work, however, we adopt a simple and robust heuristic: we set $\tau$ proportional to the relative size of the target domain with respect to the source domain (in terms of training examples), and distribute the remaining source-domain proportions equally, i.e., $\sigma_a = \sigma_{ta} = \sigma_t = \frac{1 - \tau}{3}$. This choice ensures that larger target domains naturally receive more adaptation weight, while still preserving source-domain coverage. It also allows us to systematically analyze the effect of varying $\tau$ across domains.\footnote{The exact $\tau$ values are provided in the result tables next to each domain.} By carefully mixing audio, projector-based noise, and synthetic textual noise in each batch, we enable the LLM to maintain its original transcription ability (i.e., not forgetting the alignment already learned by the projector) while adapting to the target domain in the absence of target-domain audio.

\begin{table}[!t]
\centering
\caption{Datasets used for training, validation, and testing in our experiments. The source partition comes from either DefinedAI (\textbf{B}anking, \textbf{I}nsurance, \textbf{H}ealthcare) or SlideSpeech (\textbf{L}ife, \textbf{T}alent, \textbf{E}nglish, \textbf{An}imation, \textbf{Ag}riculture, \textbf{M}usical \textbf{I}nstruments). For evaluation, each target partition is paired individually with the chosen source partition. For reference, we provide the total number of utterances (\textbf{\#Utts}) and the duration (\textbf{Hrs}) of each partition.}
\vspace{0.1cm}
\label{tab:data_summary_b}
\setlength{\tabcolsep}{4.4pt}
\renewcommand{\arraystretch}{1.35}
\footnotesize
\begin{tabular}{@{} l@{~} c@{~~} c@{~~} r@{~~~} l@{~~} r@{~~~} l@{~~} r@{~~~} l @{}}
\toprule
\multirow{2}{*}{\textbf{Dataset}} & \multirow{2}{*}{\textbf{Partition}} & \multirow{2}{*}{\textbf{Domains}} & \multicolumn{2}{c}{\textbf{Train}} & \multicolumn{2}{c}{\textbf{Validation}} & \multicolumn{2}{c}{\textbf{Test}} \\
\cmidrule(lr){4-5} \cmidrule(lr){6-7} \cmidrule(lr){8-9}
 &  &  & \textbf{\#Utts} & \textbf{Hrs} & \textbf{\#Utts} & \textbf{Hrs} & \textbf{\#Utts} & \textbf{Hrs} \\
\midrule
DefinedAI   & Source      & B/I/H   & 17{,}398 & 38{:}10 & --   & --    & --   & --    \\
SlideSpeech & Source     & L/T/E            & 34{,}682 & 34{:}59 & --   & --    & --   & --    \\
\midrule
DefinedAI   & Target  & B                        & 26{,}704 & 36{:}20 & 391  & 0{:}48 & 695  & 1{:}24 \\
DefinedAI   & Target  & I                      & 32{,}249 & 46{:}54 & 325  & 0{:}42 & 650  & 1{:}18 \\
SlideSpeech & Target  & Ag                    & 30{,}498 & 29{:}20 & 1{,}504 & 1{:}40 & 2{,}934 & 3{:}18 \\
SlideSpeech & Target  & An                      & 56{,}593 & 49{:}48 & 2{,}896 & 2{:}50 & 5{,}934 & 5{:}32 \\
SlideSpeech & Target  & MI            & 9{,}981  & 8{:}30  & 719  & 0{:}43 & 863  & 0{:}53 \\
\bottomrule
\end{tabular}
\vspace{-1.95em}
\normalsize
\end{table}

\section{Datasets Preparation}
Table~\ref{tab:data_summary_b} presents the composition of the source and target domain datasets, including their training, development, and test splits, as used in our experiments.

Our experiments use two conversational corpora, chosen both for their explicit domain grouping and for their relevance to real-world ASR applications. DefinedAI\footnote{https://defined.ai} is a proprietary corpus of manually transcribed customer–agent telephone calls (125 hours used), selected for its production-like conditions. Conversational data like this captures naturalistic speech patterns, e.g., including disfluencies, interruptions, and colloquial expressions, that are critical for evaluating ASR robustness in practical settings. To complement it, we use the open-source SlideSpeech~\cite{DBLP:conf/icassp/WangYSWZL24} dataset, a large-scale audio-visual dataset generated from online conference videos on YouTube, which contains multi-speaker conversations across 22 domains (1{,}705 videos; 1{,}000 hours; 473 hours transcribed). By working with these corpora, we focus on realistic conversational scenarios that are more challenging than scripted speech. 


From DefinedAI we prepared three partitions: a source (audio, transcripts pairs) partition covering Banking (B), Insurance (I) and Healthcare (H) domains, and two target (text-only) partitions with Banking (B) and Insurance (I) domains. 
For the SlideSpeech dataset, although it provides a challenging testbed for evaluating ASR robustness, the limited amount of data in its original development and test partitions (approximately 1 hour per domain) limits its usefulness for our experiments. To address this, we ranked the domains by perplexity and selected only those from the original SlideSpeech training set that contained sufficient examples for partitioning into training and development/test portions. Using this ranking, we constructed the source and target partitions: the source domains comprise Life, Talent, and English, while the target domains include Agriculture, Animation, and Musical Instruments (see Table \ref{tab:data_summary_b}). Note that, in our setup, the training split of the target domain contains text-only data.

\section{Experimental Setup}
\label{sec:exp}
All experiments were implemented in the SLAM-ASR framework~\cite{ma2024embarrassingly} with WavLM-Large~\cite{chen2022wavlm} as the speech encoder and Llama~3.2~3B Instruct\footnote{https://huggingface.co/meta-llama/Llama-3.2-1B-Instruct}~\cite{dubey2024llama} as the decoder LLM. These models are connected via a learnable projector consisting of a single hidden layer followed by a ReLU activation and a regression layer. For all the performed experiments, the prompt template is defined as:
\begin{description}[noitemsep,leftmargin=1.2em]
\begin{small}\ttfamily
\item \textsc{Prompt}(\textbf{input}) = \textcolor{prompt}{<|start\_header\_id|>user}
\item \textcolor{prompt}{<|end\_header\_id|>}\textbf{\textcolor{prompt}{Transcribe speech to text. Speech:}}\textbf{\textcolor{black}{\{input\}}}\textcolor{prompt}{<|eot\_id|><|start\_header\_id|>}
\item \textcolor{prompt}{assistant<|end\_header\_id|>}\textbf{\{transcript\}}
\end{small}
\end{description}

Notice that the value of \textbf{\texttt{\{input\}}} will change with the type of (input, output) pair in the batch composition, in particular:
\begin{itemize} 
\item \textsc{Prompt}($a$) for $(a, t)$
\item \textsc{Prompt}($noise_a(t)$) for $(noise_a(t), t)$
\item \textsc{Prompt}($noise(t)$) for $(noise(t), t)$,
\end{itemize}
where \textbf{\{transcript\}} is always the ground truth transcript $t$.


\noindent $\bullet$ \textbf{Base model}: We train a \emph{base model} for each dataset (DefinedAI and SlideSpeech) using the source (audio-text pairs) partition, and evaluated on the test split of the target data. For training the base model, we followed the original setup as described in~\cite{ma2024embarrassingly}: the speech encoder and the LLM are frozen, while the projector is trained on the source domain data during four epochs with learning rate $1\!\times\!10^{-4}$, $1000$ warm-up steps, and batch size of $4$. 

\noindent$\bullet$ \textbf{Adapted model (audio)}: This experiment reflects the best-case scenario, i.e., when audio-text pairs are available for fine-tuning the LLM for the target domain. Specifically, for adapting the LLM, we fine-tuned on the target partition, using audio-text pairs, for four epochs, using LoRA applied to the self-attention \emph{query} and \emph{value} projection layers with rank 8, learning rate $1\!\times\!10^{-4}$, and $1000$ warm-up steps.

\noindent$\bullet$ \textbf{Text-only adaptation (ours)}: For these experiments, we implement \emph{noise(text)} as a two-step process: (1) \emph{random character substitution} followed by (2) \emph{random character duplication}. 
In step (1), we use nlpaug\footnote{https://nlpaug.readthedocs.io/en/latest/augmenter/char/random.html} to select $15\%$ of the words and replace $30\%$ of the characters within those words with random symbols, with a minimum of 1 and a maximum of 10 character edits per utterance. These values match the nlpaug defaults, except that we reduce the fraction of edited words from $30\%$ to $15\%$, which yielded better performance in preliminary experiments. 
In step (2), each character has a $10\%$ chance of being repeated; when selected, it is duplicated 1 to 3 times with equal probability. This step is designed to emulate the duplication patterns observed in noise(audio).

Additionally, we also reimplemented two recently proposed text-only adaptation approaches for LLM-based ASR: 

\noindent$\bullet$ \emph{\textbf{Ma et al.}}~\cite{DBLP:conf/slt/MaLK24}: the text-only adaptation is performed in two stages: (1) first, freeze the model and learn $k$ trainable embeddings $e_i$ inserted where the audio would go in the prompt (i.e. the input of the LLM is \textsc{Prompt($e_1\dots e_k$)}); (2) fine-tune the LLM using those $k$ embeddings instead of audio, i.e., batches are composed only of $(e_1\dots e_k, t)$ pairs. In our experiments, we use $k=30$ as in the original work.

\noindent$\bullet$ \emph{\textbf{Fang et al.}}~\cite{DBLP:journals/corr/abs-2506-05671}: the adaptation process consists of: (a) fine-tuning the LLM using raw target-domain text (no prompt involved); (b) monitoring the perplexity on the validation set (with audio) every 200 steps to identify when catastrophic forgetting occurs. The adapted model is the checkpoint with the minimum perplexity before the sudden increase.


\begin{table}[!t]
\centering
\caption{In-domain WER results. Domain adaptation on DefinedAI (target). Base model trained on DefinedAI (source). Values in \%.}
\vspace{0.1cm}
\label{tab:results_defai_to_defai}
\scriptsize
\setlength{\tabcolsep}{9.5pt}
\begin{tabular}{@{} l cc cc @{}}
\toprule
\multirow{2}{*}{\textbf{System}}
  & \multicolumn{2}{c}{\shortstack{\textbf{Banking}\\($\boldsymbol\tau=0.61$)}}
  & \multicolumn{2}{c}{\shortstack{\textbf{Insurance}\\($\boldsymbol\tau=0.65$)}} \\
\cmidrule(lr){2-3}\cmidrule(lr){4-5}
 & \textbf{WER} & $\Delta$ & \textbf{WER} & $\Delta$ \\
\midrule
Base model                           & 12.98 & --     & 10.61 & --     \\
\midrule
\textit{Adapted model (audio)}       &  9.92 & 23.6 &  7.92 & 25.4 \\
\midrule
\textit{Adapted model (text)}        &       &        &       &        \\
\quad Fang et al.~\cite{DBLP:journals/corr/abs-2506-05671} & 10.92 & 15.9 &  9.79 &  7.7 \\
\quad Ma et al.~\cite{DBLP:conf/slt/MaLK24}               & 10.63 & 18.1 &  9.68 &  8.8 \\
\quad \textbf{Ours}                  & \textbf{10.11} & \textbf{22.1} & \textbf{8.71} & \textbf{17.9} \\
\bottomrule
\end{tabular}
\vspace{-0.35em}
\end{table}

Finally, all experiments are evaluated in terms of word error rate (WER), and report the relative improvement ($\Delta$) over the corresponding \textit{base model}. 

\section{Experimental Results}

We evaluate our proposed text-only adaptation method in three scenarios of increasing difficulty: (1) in-domain adaptation, (2) out-of-domain adaptation, (3) cross-domain adaptation. The main results for each of these experiments are reported in Table~\ref{tab:results_defai_to_defai}, Table~\ref{tab:results_ss_to_ss} and Table~\ref{tab:results_defai_to_ss}, respectively. 


\noindent\textbf{(1) In-domain adaptation} - In this setting, the target domain corresponds to a domain type already represented in the source-domain data, both in terms of domain exposure and similar speech and acoustic characteristics. The goal of this experiment was to assess the benefit of additional text data for a domain familiar to the base model. Specifically, for this experiment, the source and target partitions are both drawn from DefinedAI. 

Table~\ref{tab:results_defai_to_defai} reports the in-domain results. After text-only adaptation, performance approaches that of audio-based adaptation (best case), with 10.11\% vs. 9.92\% in Banking and 8.71\% vs. 7.92\% in Insurance, highlighting the benefits of incorporating additional text data for a familiar domain.

\noindent\textbf{(2) Out-of-domain adaptation} - In this setting, the target domain is not represented in the source domain partition but shares the same speech and acoustic characteristics. The goal of this experiment was to validate how well the LLM can learn domain-specific lexical and syntactic patterns from text alone, given stable acoustic conditions. This scenario was simulated using the SlideSpeech dataset, where source domain data is represented by L,T,E domains and the target domain is defined by Ag, An, MI domains. 

Table~\ref{tab:results_ss_to_ss} presents the out-of-domain results. Our method achieves consistent WER improvements in two of the three target domains, indicating that the LLM learns domain-specific lexicons, albeit modestly due to the low $\boldsymbol\tau$ values ($<0.25$ in MI). This suggests that higher specialization (larger $\boldsymbol\tau$) could yield further improvements.

\begin{table}[!t]
\centering
\caption{Out-of-domain WER results.  Domain adaptation on SlideSpeech (target). Base model trained on SlideSpeech (source). Values in \%.}
\vspace{0.1cm}
\label{tab:results_ss_to_ss}
\scriptsize
\setlength{\tabcolsep}{3.7pt}
\begin{tabular}{@{} l cc cc cc @{}}
\toprule
\multirow{2}{*}{\textbf{System}}
  & \multicolumn{2}{c}{\shortstack{\textbf{Ag}\\($\boldsymbol\tau=0.47$)}}
  & \multicolumn{2}{c}{\shortstack{\textbf{An}\\($\boldsymbol\tau=0.62$)}}
  & \multicolumn{2}{c}{\shortstack{\textbf{MI}\\($\boldsymbol\tau=0.22$)}} \\
\cmidrule(lr){2-3}\cmidrule(lr){4-5}\cmidrule(lr){6-7}
 & \textbf{WER} & $\Delta$ & \textbf{WER} & $\Delta$ & \textbf{WER} & $\Delta$\\
\midrule
Base model                           & 14.82 & --     & 15.58 & --      & 14.73 & --     \\
\midrule
\textit{Adapted model (audio)}       & 10.80 & 27.1 & 10.37 & 33.4  & 11.04 & 25.1 \\
\midrule
\textit{Adapted model (text)}        &       &        &       &         &       &        \\
\quad Fang et al.~\cite{DBLP:journals/corr/abs-2506-05671} & 14.47 & 2.4  & 15.30 & 1.8   & 13.70 & 7.0  \\
\quad Ma et al.~\cite{DBLP:conf/slt/MaLK24}               & 14.23 & 4.0  & 15.71 & $-0.8$ & \textbf{13.35} & \textbf{9.4}  \\
\quad \textbf{Ours}                                       & \textbf{14.21} & \textbf{4.1} & \textbf{14.60} & \textbf{6.3} & 13.43 & 8.8 \\
\bottomrule
\end{tabular}
\vspace{-1.95em}
\end{table}

\noindent\textbf{(3) Cross-domain adaptation} - This setting is the most challenging and realistic scenario; the target domain is not represented in the source data and additionally exhibits different speech and acoustic characteristics. The goal of this experiment was to evaluate the impact of the text-only adaptation approach in bridging the linguistic gap in a scenario where there are evident acoustic shifts. For this experiment, we use DefinedAI (B/I/H) as the source domain and SlideSpeech (Ag/Ai/MI) as the target domain.

Table~\ref{tab:results_defai_to_ss} shows that our text-only adaptation approach improves over the base model, achieving performance comparable to Ma et al.~\cite{DBLP:conf/slt/MaLK24}. This indicates that the method can reduce the linguistic gap between domains that differ in both lexicon and acoustics.
However, performance remains well below that of the audio-adapted model, an expected outcome given that the latter benefits from target-domain audio to address the acoustic mismatch.

\begin{table}[t]
\centering
\caption{Cross-domain WER results. Domain adaptation on SlideSpeech (target). Base model trained on DefinedAI (source). Values in \%.}
\vspace{0.1cm}
\label{tab:results_defai_to_ss}
\scriptsize
\setlength{\tabcolsep}{3.7pt}
\begin{tabular}{@{} l cc cc cc @{}}
\toprule
\multirow{2}{*}{\textbf{System}}
  & \multicolumn{2}{c}{\shortstack{\textbf{Ag}\\($\boldsymbol\tau=0.64$)}}
  & \multicolumn{2}{c}{\shortstack{\textbf{An}\\($\boldsymbol\tau=0.77$)}}
  & \multicolumn{2}{c}{\shortstack{\textbf{MI}\\($\boldsymbol\tau=0.37$)}} \\
\cmidrule(lr){2-3}\cmidrule(lr){4-5}\cmidrule(lr){6-7}
 & \textbf{WER} & $\Delta$ & \textbf{WER} & $\Delta$ & \textbf{WER} & $\Delta$ \\
\midrule
Base model                           & 32.64 & --     & 29.81 & --     & 28.00 & --     \\
\midrule
\textit{Adapted model (audio)}       & 12.25 & 62.5 & 11.23 & 62.3 & 13.20 & 52.9 \\
\midrule
\textit{Adapted model (text)}        &       &        &       &        &       &        \\
\quad Fang et al.~\cite{DBLP:journals/corr/abs-2506-05671} & 31.01 &  5.0 & 27.83 &  6.6 & 26.72 &  4.6 \\
\quad Ma et al.~\cite{DBLP:conf/slt/MaLK24}               & 29.22 & 10.5 & 25.95 & 12.9 & 25.03 & 10.6 \\
\quad \textbf{Ours}                                       & \textbf{29.18} & \textbf{10.6} & \textbf{25.32} & \textbf{15.1} & \textbf{23.54} & \textbf{15.9} \\
\bottomrule
\end{tabular}
\vspace{-1.95em}
\end{table}


\subsection{Ablation experiments}
We conducted two ablation studies on the DefinedAI target data to isolate the contributions of our method's key components.
The first study evaluates the impact of including or removing different source-side components from the training batches. As shown in Table~\ref{tab:abl_interleave_wide_bt}, using all three components ($\sigma_a$, $\sigma_{ta}$, $\sigma_t$) yields the best performance.
Notably, omitting the audio component ($\sigma_a = 0$) causes a sharp increase in WER, possibly due to catastrophic forgetting of the speech-text alignment. The second study examines the importance of perturbing text-only data from the target domain. Table~\ref{tab:ablation_definedai} shows that using syntactic noise as input to the LLM improves performance more effectively than using unperturbed text. This indicates that framing the task as denoising enables the model to better capture the target domain's lexical and syntactic patterns.

\begin{table}[t]
\centering
\caption{Batch composition ablation study. Checkmarks denote the active components in the batch. $\Delta$ is reported over the base model of Table~\ref{tab:results_defai_to_defai}. Values in \%.}
\vspace{0.1cm}
\label{tab:abl_interleave_wide_bt}
\scriptsize
\setlength{\tabcolsep}{8pt}
\begin{tabular}{@{} ccc | S[table-format=2.2] c S[table-format=2.2] c @{}}
\toprule
\multirow{2}{*}{$\sigma_a$} & \multirow{2}{*}{$\sigma_{ta}$} & \multirow{2}{*}{$\sigma_t$} & \multicolumn{2}{c}{\textbf{Banking}} & \multicolumn{2}{c}{\textbf{Insurance}} \\
\cmidrule(lr){4-5}\cmidrule(lr){6-7}
 & & & \textbf{WER} & {$\Delta$} & \textbf{WER} & {$\Delta$} \\
\midrule
\checkmark &            &            & 11.29 & 13.0  &  9.91 &  6.6 \\
\checkmark & \checkmark &            & 10.14 & 21.9  &  8.94 & 15.7 \\
\checkmark &            & \checkmark & 11.77 &  9.3  &  9.10 & 14.2 \\
           & \checkmark & \checkmark & 73.50 & -466.3 & 66.65 & -528.2 \\
\checkmark & \checkmark & \checkmark & \textbf{10.11} & \textbf{22.1} & \textbf{8.71} & \textbf{17.9} \\
\bottomrule
\end{tabular}
\vspace{-1.95em}
\end{table}

\begin{table}[!htpb]
\centering
\scriptsize
\caption{Batch composition with respect to the LLM input format for $\sigma_t$ and $\tau$. The original noise is replaced by different strategies. $\Delta$ is reported over the base model of Table~\ref{tab:results_defai_to_defai}. Values in \%.}
\vspace{0.1cm}
\label{tab:ablation_definedai}
\setlength{\tabcolsep}{2.2pt}
\renewcommand{\arraystretch}{1.00}
\begin{tabular}{p{2cm} p{2.5cm} S[table-format=2.2] c S[table-format=2.2] c}
\toprule
\multirow{2}{*}{\textbf{Strategy}} &\multirow{2}{*}{\textbf{Input}} & \multicolumn{2}{c}{\textbf{Banking}} & \multicolumn{2}{c}{\textbf{Insurance}} \\
\cmidrule(lr){3-4}\cmidrule(lr){5-6}
& & \textbf{WER} & {$\Delta$} & \textbf{WER} & {$\Delta$} \\
\midrule
Noise & \texttt{\textsc{Prompt}}($noise(t)$) & \bfseries 10.11 & \bfseries 22.1 & \bfseries 8.71 & \bfseries 17.9 \\
\midrule
Echo & \texttt{\textsc{Prompt}}($t$)        & 10.31 & 20.6 & 9.09 & 14.3 \\
Empty prompt & \texttt{\textsc{Prompt}}()                                 & 10.56 & 18.6 & 9.01 & 15.1 \\

No prompt & $t$                                       & 10.53 & 18.9 & 9.28 & 12.5 \\
\bottomrule
\end{tabular}
\vspace{-1.95em}
\end{table}


\section{Conclusions}

This paper presented a novel text-only adaptation approach for LLM-based ASR architectures that mitigates catastrophic forgetting when fine-tuning with textual data alone. Our method integrates multiple components during fine-tuning—alternating between source audio, projector-induced text noise, and synthetic noisy text—allowing the model to simultaneously (i) preserve its ability to interpret audio-conditioned inputs and (ii) learn to denoise text, while also acquiring target-domain knowledge. Experiments demonstrate consistent improvements across multiple domains and datasets, surpassing prior state-of-the-art text-only adaptation strategies. As future work, we plan to explore more sophisticated \textit{noise} functions that better approximate the projection layer outputs and to conduct a thorough evaluation of the $\tau$ value to assess performance under text-rich, real-world conditions. 




\bibliographystyle{IEEEbib}
\bibliography{paper}

@article{DBLP:journals/corr/abs-2506-05671,
  author       = {Yangui Fang and
                  Jing Peng and
                  Xu Li and
                  Yu Xi and
                  Chengwei Zhang and
                  Guohui Zhong and
                  Kai Yu},
  title        = {Low-Resource Domain Adaptation for Speech {LLMs} via Text-Only Fine-Tuning},
  journal      = {CoRR},
  volume       = {abs/2506.05671},
  year         = {2025},
  doi          = {10.48550/ARXIV.2506.05671},
  eprinttype    = {arXiv},
  eprint       = {2506.05671},
}

@inproceedings{DBLP:conf/slt/MaLK24,
  author       = {Yingyi Ma and
                  Zhe Liu and
                  Ozlem Kalinli},
  title        = {Effective Text Adaptation For LLM-Based {ASR} Through Soft Prompt Fine-Tuning},
  booktitle    = {{IEEE} Spoken Language Technology Workshop, {SLT} 2024, Macao, December
                  2-5, 2024},
  pages        = {64--69},
  publisher    = {{IEEE}},
  year         = {2024},
  doi          = {10.1109/SLT61566.2024.10832227},
}

@article{chen2022wavlm,
  title={{WavLM}: Large-scale self-supervised pre-training for full stack speech processing},
  author={Chen, Sanyuan and Wang, Chengyi and Chen, Zhengyang and Wu, Yu and Liu, Shujie and Chen, Zhuo and Li, Jinyu and Kanda, Naoyuki and Yoshioka, Takuya and Xiao, Xiong and others},
  journal={{IEEE} Journal of Selected Topics in Signal Processing},
  volume={16},
  number={6},
  pages={1505--1518},
  year={2022},
  publisher={IEEE},
  doi= {10.1109/JSTSP.2022.3188113},
}

@article{dubey2024llama,
  title={The {Llama} 3 Herd of Models},
  author={Dubey, Abhimanyu and Jauhri, Abhinav and Pandey, Abhinav and Kadian, Abhishek and Al-Dahle, Ahmad and Letman, Aiesha and Mathur, Akhil and Schelten, Alan and Yang, Amy and Fan, Angela and others},
  journal      = {CoRR},
  volume       = {abs/2407.21783},
  year         = {2024},
  doi          = {10.48550/ARXIV.2407.21783},
  eprinttype    = {arXiv}
}

@inproceedings{DBLP:conf/icassp/WangYSWZL24,
  author       = {Haoxu Wang and
                  Fan Yu and
                  Xian Shi and
                  Yuezhang Wang and
                  Shiliang Zhang and
                  Ming Li},
  title        = {SlideSpeech: {A} Large Scale Slide-Enriched Audio-Visual Corpus},
  booktitle    = {ICASSP, Seoul, Republic of Korea},
  pages        = {11076--11080},
  publisher    = {{IEEE}},
  year         = {2024},
  doi          = {10.1109/ICASSP48485.2024.10448079}
}

@article{KHEDDAR2024102422,
title = {Automatic speech recognition using advanced deep learning approaches: A survey},
journal = {Information Fusion},
volume = {109},
pages = {102422},
year = {2024},
issn = {1566-2535},
doi = {https://doi.org/10.1016/j.inffus.2024.102422},
url = {https://www.sciencedirect.com/science/article/pii/S1566253524002008},
author = {Hamza Kheddar and Mustapha Hemis and Yassine Himeur}
}

@inproceedings{zhang2023speechgpt,
  title={{SpeechGPT}: Empowering Large Language Models with Intrinsic Cross-Modal Conversational Abilities},
  author={Zhang, Dong and Li, Shimin and Zhang, Xin and Zhan, Jun and Wang, Pengyu and Zhou, Yaqian and Qiu, Xipeng},
  booktitle={Findings of the Association for Computational Linguistics: EMNLP 2023},
  pages={15757--15773},
  year={2023}
}

@article{goel2025audio,
  title={{Audio Flamingo} 3: Advancing Audio Intelligence with Fully Open Large Audio Language Models},
  author={Goel, Arushi and Ghosh, Sreyan and Kim, Jaehyeon and Kumar, Sonal and Kong, Zhifeng and Lee, Sang-gil and Yang, Chao-Han Huck and Duraiswami, Ramani and Manocha, Dinesh and Valle, Rafael and others},
  journal={arXiv preprint arXiv:2507.08128},
  year={2025}
}

@article{das2024speechverse,
  title={{SpeechVerse}: A large-scale generalizable audio language model},
  author={Das, Nilaksh and Dingliwal, Saket and Ronanki, Srikanth and Paturi, Rohit and Huang, Zhaocheng and Mathur, Prashant and Yuan, Jie and Bekal, Dhanush and Niu, Xing and Jayanthi, Sai Muralidhar and others},
  journal={arXiv preprint arXiv:2405.08295},
  year={2024}
}

@article{chu2024qwen2,
  title={Qwen2-audio technical report},
  author={Chu, Yunfei and Xu, Jin and Yang, Qian and Wei, Haojie and Wei, Xipin and Guo, Zhifang and Leng, Yichong and Lv, Yuanjun and He, Jinzheng and Lin, Junyang and others},
  journal={arXiv preprint arXiv:2407.10759},
  year={2024}
}

@article{ma2024embarrassingly,
  title={An Embarrassingly Simple Approach for {LLM} with Strong {ASR} Capacity},
  author={Ma, Ziyang and Yang, Guanrou and Yang, Yifan and Gao, Zhifu and Wang, Jiaming and Du, Zhihao and Yu, Fan and Chen, Qian and Zheng, Siqi and Zhang, Shiliang and others},
  journal={arXiv preprint arXiv:2402.08846},
  year={2024}
}

@INPROCEEDINGS{10445874,
  author={Yu, Wenyi and Tang, Changli and Sun, Guangzhi and Chen, Xianzhao and Tan, Tian and Li, Wei and Lu, Lu and Ma, Zejun and Zhang, Chao},
  booktitle={ICASSP}, 
  title={Connecting Speech Encoder and Large Language Model for {ASR}}, 
  year={2024},
  volume={},
  number={},
  pages={12637-12641},
  doi={10.1109/ICASSP48485.2024.10445874}}

@INPROCEEDINGS{kumar2024performance,
  author={Kumar, Shashi and Thorbecke, Iuliia and Burdisso, Sergio and Villatoro-Tello, Esaú and K E, Manjunath and Hacioğlu, Kadri and Rangappa, Pradeep and Motlicek, Petr and Ganapathiraju, Aravind and Stolcke, Andreas},
  booktitle={ICASSPW}, 
  title={Performance Evaluation of {SLAM-ASR}: The Good, the Bad, the Ugly, and the Way Forward}, 
  year={2025},
  volume={},
  number={},
  pages={1-5},
  doi={10.1109/ICASSPW65056.2025.11010998}}

@misc{yang2025bridgingmodalitygapsoftly,
      title={Bridging the Modality Gap: Softly Discretizing Audio Representation for {LLM}-based Automatic Speech Recognition}, 
      author={Mu Yang and Szu-Jui Chen and Jiamin Xie and John Hansen},
      year={2025},
      eprint={2506.05706},
      archivePrefix={arXiv},
      primaryClass={eess.AS},
      url={https://arxiv.org/abs/2506.05706}, 
}

@article{sedlavcek2025approaching,
  title={Approaching Dialogue State Tracking via Aligning Speech Encoders and {LLMs}},
  author={Sedl{\'a}{\v{c}}ek, {\v{S}}imon and Yusuf, Bolaji and {\v{S}}vec, J{\'a}n and Hegde, Pradyoth and Kesiraju, Santosh and Plchot, Old{\v{r}}ich and {\v{C}}ernock{\`y}, Jan},
  journal={arXiv preprint arXiv:2506.08633},
  year={2025}
}

@inproceedings{wang2023slm,
  title={{SLM}: Bridge the thin gap between speech and text foundation models},
  author={Wang, Mingqiu and Han, Wei and Shafran, Izhak and Wu, Zelin and Chiu, Chung-Cheng and Cao, Yuan and Chen, Nanxin and Zhang, Yu and Soltau, Hagen and Rubenstein, Paul K and others},
  booktitle={ASRU Proceedings},
  pages={1--8},
  year={2023},
  organization={IEEE}
}

@inproceedings{yang24f_interspeech,
  title     = {MaLa-{ASR}: Multimedia-Assisted {LLM}-Based ASR},
  author    = {Guanrou Yang and Ziyang Ma and Fan Yu and Zhifu Gao and Shiliang Zhang and Xie Chen},
  year      = {2024},
  booktitle = {Interspeech 2024},
  pages     = {2405--2409},
  doi       = {10.21437/Interspeech.2024-488},
  issn      = {2958-1796},
}

@inproceedings{radford2023robust,
  title={Robust speech recognition via large-scale weak supervision},
  author={Radford, Alec and Kim, Jong Wook and Xu, Tao and Brockman, Greg and McLeavey, Christine and Sutskever, Ilya},
  booktitle={International conference on machine learning},
  pages={28492--28518},
  year={2023},
  organization={PMLR}
}

@article{hsu2021hubert,
  title={Hubert: Self-supervised speech representation learning by masked prediction of hidden units},
  author={Hsu, Wei-Ning and Bolte, Benjamin and Tsai, Yao-Hung Hubert and Lakhotia, Kushal and Salakhutdinov, Ruslan and Mohamed, Abdelrahman},
  journal={IEEE/ACM transactions on audio, speech, and language processing},
  volume={29},
  pages={3451--3460},
  year={2021},
  publisher={IEEE}
}

@INPROCEEDINGS{10800077,
    author={Geng, Xuelong and Xu, Tianyi and Wei, Kun and Mu, Bingshen and Xue, Hongfei and Wang, He and Li, Yangze and Guo, Pengcheng and Dai, Yuhang and Li, Longhao and Shao, Mingchen and Xie, Lei},
  booktitle={ISCSLP}, 
  title={Unveiling the Potential of {LLM}-Based {ASR} on Chinese Open-Source Datasets}, 
  year={2024},
  volume={},
  number={},
  pages={26-30},
  doi={10.1109/ISCSLP63861.2024.10800077}}

@misc{Rajaa_SpeechLLM_Multi-Modal_LLM,
author = {Rajaa, Shangeth and Tushar, Abhinav},
title = {{SpeechLLM: Multi-Modal LLM for Speech Understanding}},
howpublished = {https://github.com/skit-ai/SpeechLLM},
year = {2024},
}

@misc{vicuna2023,
    title = {Vicuna: An Open-Source Chatbot Impressing GPT-4 with 90\%* ChatGPT Quality},
    url = {https://lmsys.org/blog/2023-03-30-vicuna/},
    author = {Chiang, Wei-Lin and Li, Zhuohan and Lin, Zi and Sheng, Ying and Wu, Zhanghao and Zhang, Hao and Zheng, Lianmin and Zhuang, Siyuan and Zhuang, Yonghao and Gonzalez, Joseph E. and Stoica, Ion and Xing, Eric P.},
    month = {March},
    year = {2023}
}

@inproceedings{huang2024multilingual,
  title={{Multilingual and Fully Non-Autoregressive {ASR} with Large Language Model Fusion: A Comprehensive Study}},
  author={Huang, W Ronny and Allauzen, Cyril and Chen, Tongzhou and Gupta, Kilol and Hu, Ke and Qin, James and Zhang, Yu and Wang, Yongqiang and Chang, Shuo-Yiin and Sainath, Tara N},
  booktitle={ICASSP},
  year={2024},
  organization={IEEE}
}

@inproceedings{li2023prompting,
  title={{Prompting Large Language Models for Zero-Shot Domain Adaptation in Speech Recognition}},
  author={Li, Yuang and Wu, Yu and Li, Jinyu and Liu, Shujie},
  booktitle={ASRU Proceedings},
  pages={1--8},
  year={2023},
  organization={IEEE}
}

@article{ma2023can,
  title={{Can Generative Large Language Models Perform {ASR} Error Correction?}},
  author={Ma, Rao and Qian, Mengjie and Manakul, Potsawee and Gales, Mark and Knill, Kate},
  journal={arXiv preprint arXiv:2307.04172},
  year={2023}
}

@inproceedings{yang2023generative,
  title={{Generative Speech Recognition Error Correction With Large Language Models and Task-Activating Prompting}},
  author={Yang, Chao-Han Huck and Gu, Yile and Liu, Yi-Chieh and Ghosh, Shalini and Bulyko, Ivan and Stolcke, Andreas},
  booktitle={ASRU},
  pages={1--8},
  year={2023},
}

\end{document}